%% file: pp_SearchForJetQuenching.tex
\documentclass[3p,times,procedia]{elsarticle}
\usepackage{nupha_ecrc}

\volume{00}

\firstpage{1}

\journalname{Nuclear Physics A}

\runauth{P. M. Jacobs et al.}

\jid{nupha}

\jnltitlelogo{Nuclear Physics A}

\usepackage{amssymb}

\usepackage[figuresright]{rotating}

\input{Include}

\begin{document}

\begin{frontmatter}



\dochead{XXVIIIth International Conference on Ultrarelativistic Nucleus-Nucleus Collisions\\ (Quark Matter 2019)}

\title{Search for jet quenching effects in high multiplicity \pp\ collisions at $\sqrts=13$\, TeV}


\author{P. M. Jacobs for the ALICE Collaboration}


\address{Lawrence Berkeley National Laboratory}

\begin{abstract}
The ALICE Collaboration reports a search for jet quenching effects in \pp\ collisions at $\sqrts=13$\, TeV, in events selected on high multiplicity compared to the minimum bias population. The measurement is based on the semi-inclusive acoplanarity distribution of jets recoiling from a high-\pT\ trigger hadron. Significant broadening of the recoil jet acoplanarity distribution is observed in high multiplicity \pp\ collisions, in both data and in simulations based on the PYTHIA model. Analysis is ongoing to elucidate the origin of this effect.
\end{abstract}




\end{frontmatter}


\section{Introduction}
\label{Sect:Intro}

High-energy collisions of small systems, in which one or both 
projectiles is a proton or light nucleus, exhibit evidence of 
collective flow~\cite{Khachatryan:2013ppb,Abelev:2013ppb,PHENIX:2018lia}, 
raising the question whether a Quark-Gluon Plasma (QGP) is produced in such 
collisions~\cite{Weller:2017tsr,Mace:2018yvl}. Jet 
quenching, the modification of a jet shower due to its
interaction with dense QCD matter, is a necessary consequence of the formation 
of a QGP. Observation of jet quenching effects in small collision systems 
would provide compelling evidence of the formation of a QGP. 

Jet quenching generates multiple related phenomena: yield 
suppression of high-\pT\ hadrons and of jets with finite radius \rr; 
modification of jet substructure; and jet centroid deflection (medium-induced 
acoplanarity). Experimentally, observables based on these various phenomena have 
different sensitivities to jet quenching effects, depending upon collision 
system and kinematics. 

The search for jet quenching effects in small systems is carried 
out by comparing distributions from collisions with small and large event 
activity (EA), measured by forward multiplicity or \ET.
Inclusive hadron and jet production measurements as a function of EA in \pA\ 
collisions yield no significant evidence of yield suppression due to 
quenching 
(e.g.~\cite{Adam:2016ppb}). However, such inclusive yield 
measurements have 
limited precision due to systematic uncertainties in calculating Glauber 
scaling factors in small systems~\cite{Adam:2014qja}, and coincidence 
observables are required for more precise quenching 
measurements in small systems. The first such 
search set a limit of 400 MeV (90\% CL) for medium-induced out-of-cone energy 
transport in high vs low EA \pPb\ collisions at $\sqrtsNN=5.02$\, 
TeV~\cite{Acharya:2017okq}. Further progress requires yet better precision for 
measurements such as these, together with measurements based on new 
observables and additional small collision systems.

In these proceedings, we present the first search for jet quenching in \pp\ 
collisions. Since Glauber scaling between \pp\ event populations with 
different EA is not defined, jet quenching measurements based on inclusive yield 
suppression are not possible. In this analysis we focus on medium-induced 
acoplanarity~\cite{Appel:1985dq,Blaizot:1986ma}, which is measured via the 
semi-inclusive distribution of jets recoiling from a high-\pT\ charged hadron 
\cite{Adam:2015pbpb,Adamczyk:2017yhe}. There is significant acoplanarity even 
in the absence of in-medium broadening~\cite{Chen:2016vem}, and medium-induced 
acoplanarity is identified by additional broadening relative to this vacuum 
effect. The search for jet quenching using acoplanarity is carried out by comparing 
populations with different EA, high multiplicity (HM) vs. minimum bias (MB). 

\section{Analysis}
\label{Sect:Analysis}

Jets are reconstructed from charged tracks in the ALICE acceptance, using 
the \antikT\ algorithm with $\rr=0.4$. We utilize the observable \Drecoil\ \cite{Adam:2015pbpb}, which is the difference of two semi-inclusive distributions with two widely different ranges of trigger hadron kinematics \pTtrig\ (TT),
    
\begin{equation}
\Drecoil\left(\pTreco,\dphi\right) =  
\frac{1}{\NtrigX\ }
\frac{{\mathrm d}^{2}\Njets\ }{ {\mathrm d}\pTreco{\mathrm d}\dphi 
}\bigg|_{\pTtrig\ \in \TTSig\ }
- \cRef\ \cdot 
\frac{1}{\NtrigX\ }\frac{{\mathrm d}^{2}\Njets\ }{ {\mathrm d}\pTreco{\mathrm 
d}\dphi }\bigg|_{\pTtrig\ \in \TTRef\ },
\label{eq:Drecoil}
\end{equation}   

\noindent
where \dphi\ is the azimuthal angle between trigger hadron and recoil jet. 
The trigger hadron selections are \TTSig: $20<\pTtrig<30$ \gev\ (denoted \TT{20,30}); and 
\TTRef: $6<\pTtrig<7$ \gev\ (denoted \TT{6,7}). The standard area-based approach is applied to adjust \pTjetch\ for the median underlying-event density (\pTreco)~\cite{Cacciari:2007fd}. The constant \cRef\ is 
extracted from data \cite{Adam:2015pbpb}, with value $\sim0.95-1.0$. The \Drecoil\ observable 
provides fully data-driven suppression of background jet yield that is uncorrelated with the hadron trigger. This is of particular importance for acoplanarity measurements, in order to suppress the effects of high-$Q^2$ Multiple Parton Interactions (MPIs) (see \cite{Adam:2015pbpb} for discussion).

The data were recorded by ALICE during LHC Run 2 with \pp\ 
collisions at $\sqrts=13$\, TeV. Two online triggers were applied: an
MB trigger collected 3.2B events (equivalent to 0.098 pb$^{-1}$), and a 
high-multiplicity trigger (HM--Online), based on the forward scintillators V0A 
($2.8<\eta<5.1$) and V0C ($-3.7<\eta<-1.7$), sampled 13 pb$^{-1}$. Events are 
classified according to their summed V0 signal, $\rm{V0M}=\rm{V0A}+\rm{V0C}$. In order to combine 
runs with different V0 gain and to facilitate comparison to theoretical 
calculations, we utilize the scaled multiplicity \VzeroMscaled, where the denominator is the mean of V0M for the MB population.

\begin{figure}[tbh!]
\centering
\includegraphics[width=0.45\textwidth]{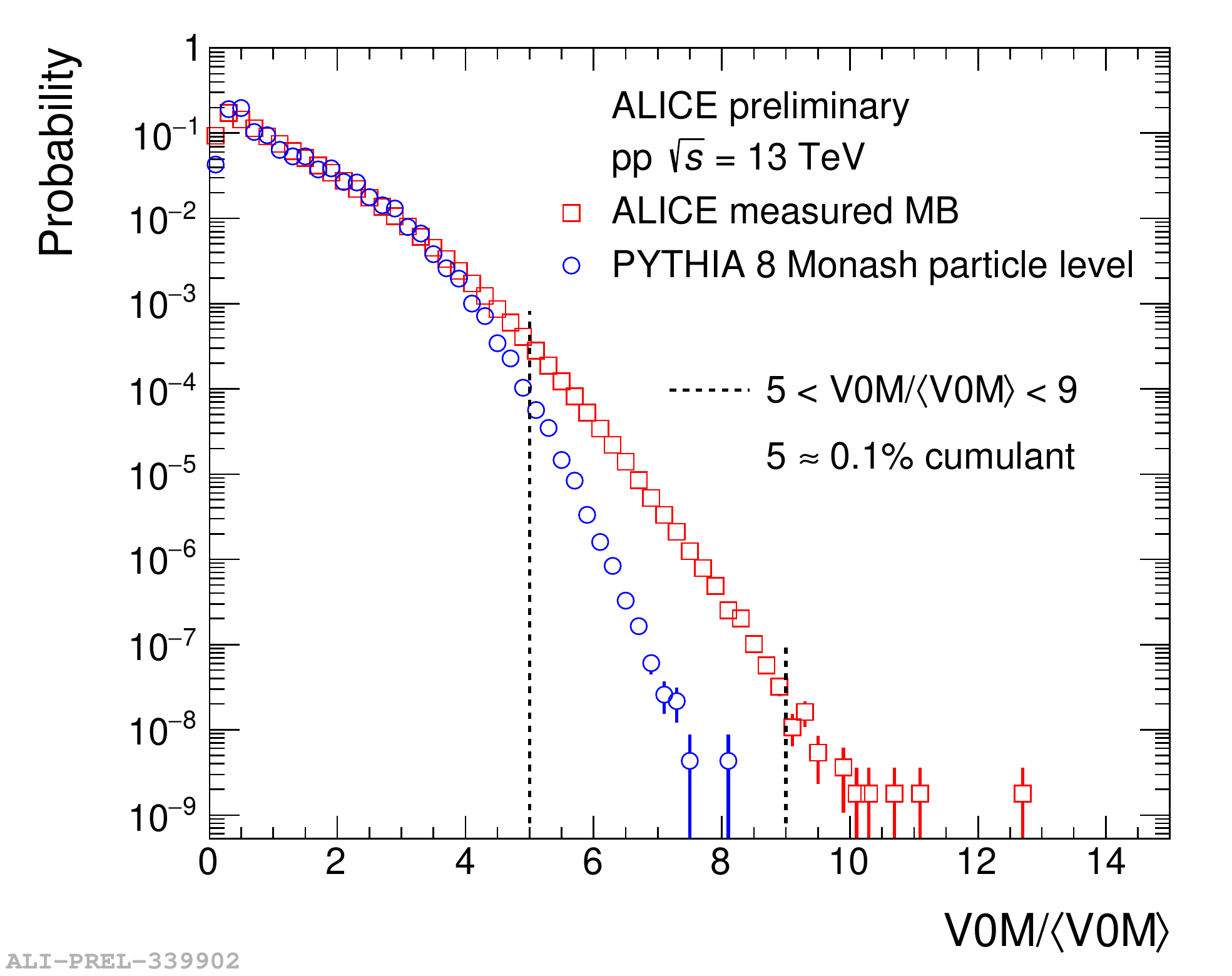}
\caption{Scaled multiplicity distribution recorded by the MB trigger. Vertical 
dashed lines indicate the High Multiplicity (HM) offline selection.}
\label{fig:ScaledV0M}
\end{figure}

Figure~\ref{fig:ScaledV0M} shows the uncorrected \VzeroMscaled\ distribution 
measured by the MB trigger (HM--Online distribution not shown). In this analysis 
we compare the MB population to an offline-selected HM population defined by 
$5<(\VzeroMscaled)<9$, indicated by the vertical dashed lines. The 
HM selection corresponds to 0.1\% of the MB cross section. The figure also shows
the \VzeroMscaled\ distribution calculated by PYTHIA Monash at the 
particle level, showing qualitative agreement with data (albeit at the 
particle level) for $(\VzeroMscaled)<3$, but lower yield than data at higher \VzeroMscaled.

\vspace*{-2.0cm}
\begin{figure}[tbh!]
\centering
\includegraphics[width=0.45\textwidth]{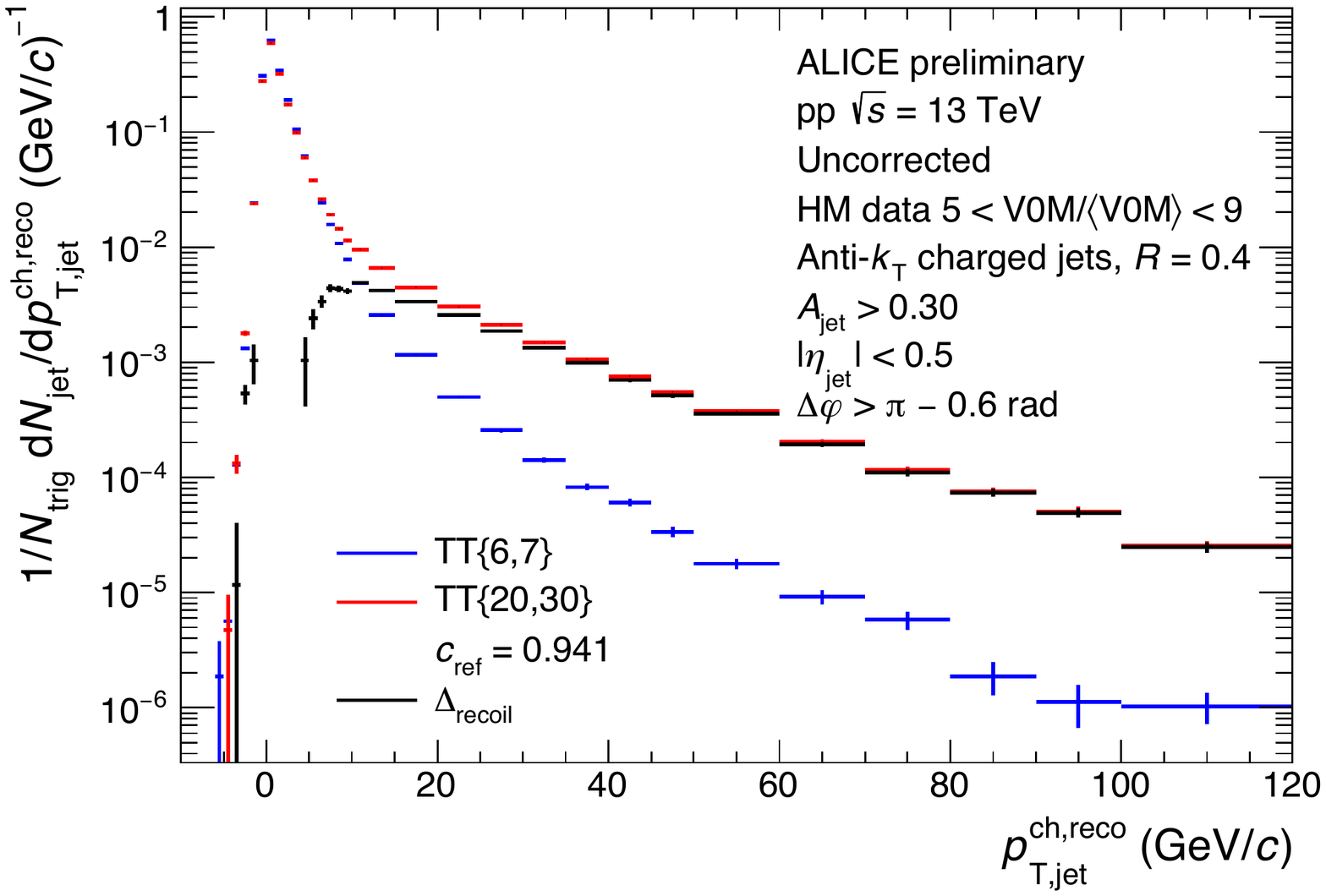}
\includegraphics[width=0.45\textwidth]{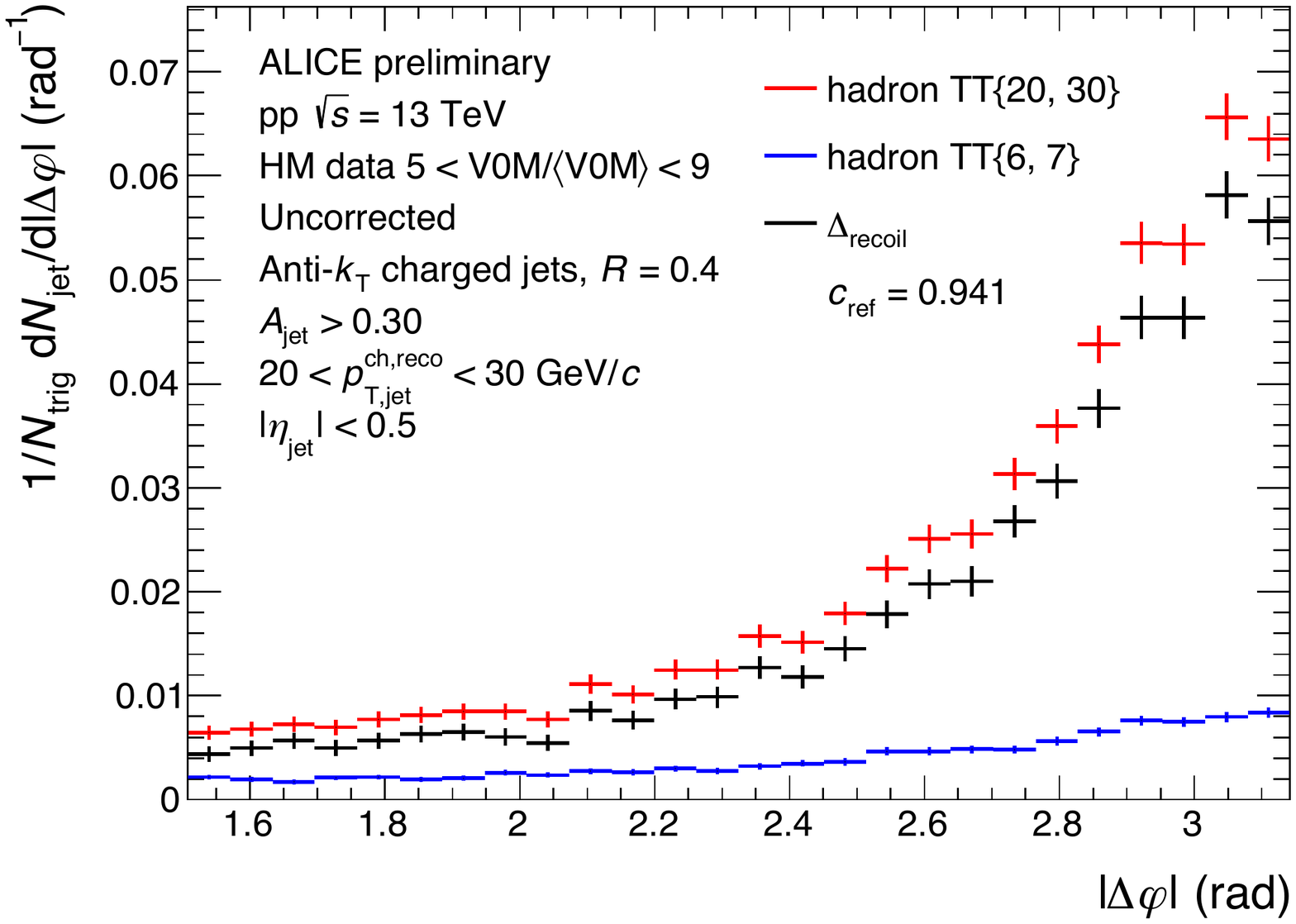}
\vspace*{-2.0cm}
\caption{Semi-inclusive distributions of recoil jets in HM-selected events for \TT{6,7}, \TT{20,30}, and the distribution of \Drecoil\ (Eq.~\ref{eq:Drecoil}). Left: vs. \pTreco\ for $\dphi>\pi-0.6$; right: vs. \dphi\ for 
$20<\pTreco<30$ \gev.}
\label{fig:Drecoil}
\end{figure}

Figure~\ref{fig:Drecoil} shows the distributions of \Drecoil\ and its two 
components as a function of \pTreco\ and \dphi. There is significant 
correlated yield at large recoil angles relative to $\dphi=\pi$, following the 
subtraction in Eq.~\ref{eq:Drecoil}.

\section{Results}
\label{Sect:Results}

\begin{figure}[tbh!]
\centering
\includegraphics[width=0.6\textwidth]{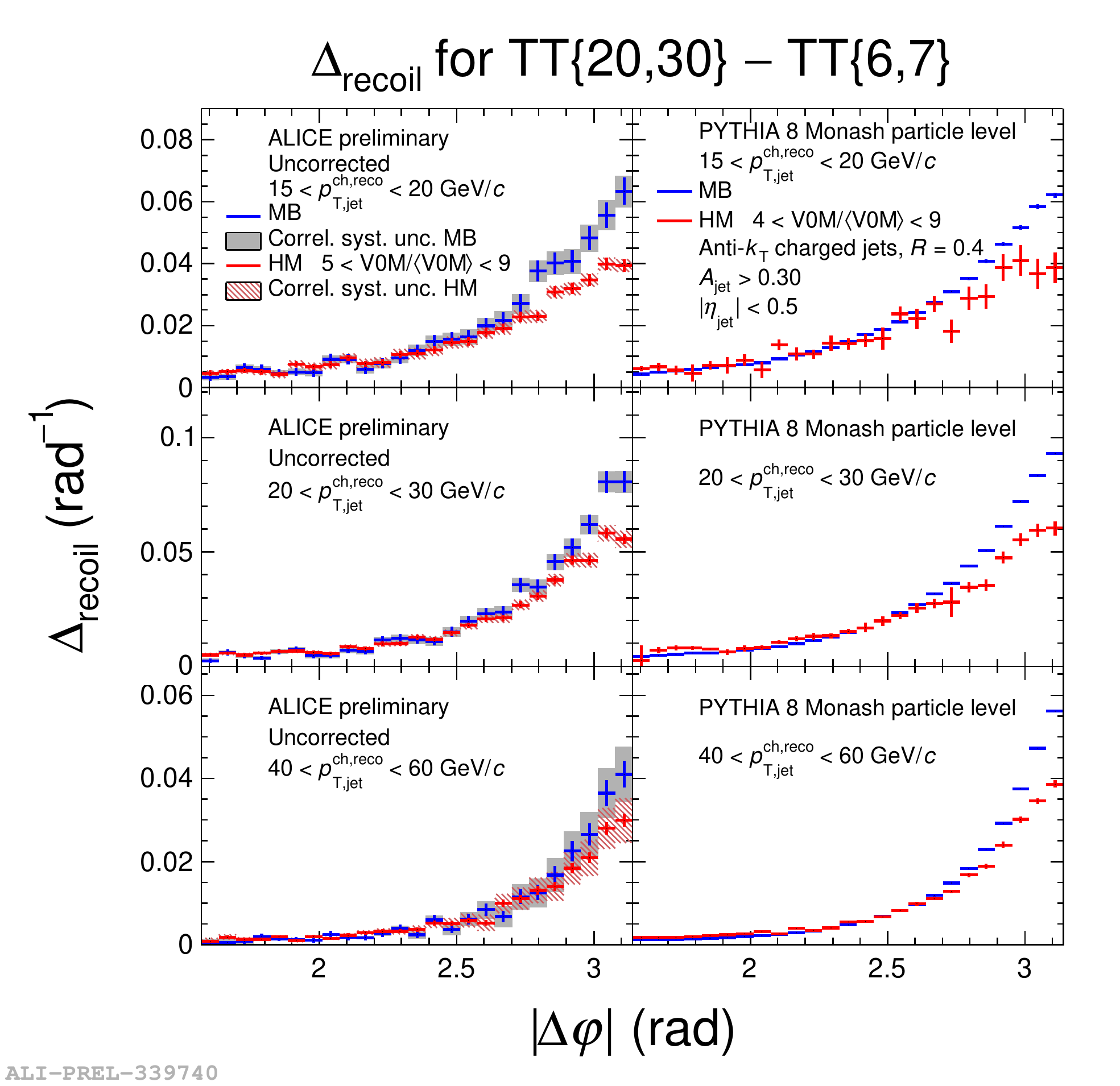}
\caption{Uncorrected acoplanarity distributions for intervals in \pTreco, for MB 
and HM selections in EA. Left: ALICE data; right: PYTHIA Monash particle-level simulation.}
\label{fig:Acoplanarity}
\end{figure}

Figure~\ref{fig:Acoplanarity}, left panels, show \dphi\ distributions in 
intervals of 
\pTreco\ for both MB and HM-selected ALICE data, not 
corrected for instrumental and background smearing effects. They exhibit a 
striking effect: for $\pTreco<40$ \gev, the HM  acoplanarity distributions are 
significantly suppressed at $\dphi\sim\pi$ and 
broadened away from the peak. This is qualitatively what is expected from jet quenching in high-EA 
collisions. 

Before concluding that jet quenching has been observed in HM \pp\ collisions, all 
other potential sources must be excluded. Azimuthal broadening due to
uncorrelated background was measured by embedding PYTHIA Monash events 
(detector-level) in HM-selected real events and 
repeating the analysis; negligible broadening is observed, so that the effect is 
not due to high track density. Likewise, the effect is observed qualitatively 
when comparing only the Signal distributions in Eq.~\ref{eq:Drecoil}, so that 
the effect is not due to the subtraction in \Drecoil. The broadening is a 
significant 
physical effect.

Figure~\ref{fig:Acoplanarity}, right panels, show similar distributions 
calculated using PYTHIA Monash at the particle level, in this case with HM corresponding 
to $4<(\VzeroMscaled)<9$; qualitative comparison only may be made to data. 
Nevertheless, the PYTHIA-generated distributions are qualitatively similar to 
those of the data, with broadened acoplanarity for HM-selected events. 

Finally, we have explored the evolution of the properties of EA-selected \pp\ 
collisions by measuring the distribution of forward-backward multiplicity 
asymmetry as a function of \VzeroMscaled\ (figures not shown). A narrowing of 
the asymmetry distribution is observed as EA increases; scaling of its width 
with EA is consistent with counting statistics being the mechanism driving the change in width, 
with clustering (e.g. due to enhanced multi-jet production at high EA) playing a lesser 
or negligible role.

\section{Outlook}
\label{Sect:Outlook}

A significant broadening of the recoil jet acoplanarity distribution in \pp\ 
collisions at $\sqrts=13$\, TeV has been observed in High EA-selected events 
compared to the MB population, for $\pTreco<40$ \gev. While this effect is 
characteristic of jet quenching, qualitatively similar features are 
observed in PYTHIA simulations. Additionally, fluctuation properties of the 
multiplicity distribution do not change significantly as a function of EA, 
suggesting similar production mechanisms at low and high-EA.
Determination of the origin of the broadening for high-EA requires new
calculations based on PYTHIA and other Monte Carlo generators, in which 
candidate mechanisms are selectively modified, with additional constraints 
imposed by measurements of forward/backward multiplicity fluctuations. Such 
calculations will help elucidate the role 
of jet quenching in the observed acoplanarity broadening.








\bibliographystyle{elsarticle-num}
\bibliography{references.bib}







\end{document}

%% file: Include.tex
\newcommand{\CommentBlock}[1]{}

\newcommand{\VzeroMscaled}{\ensuremath{\mathrm{V0M}/\left<\mathrm{V0M}\right>}}

\newcommand{\pPb}{p--Pb}

\newcommand{\pp}{pp}
\newcommand{\pA}{p--A}

\newcommand{\sqrtsNN}{\ensuremath{\sqrt{\mathrm{s}_\mathrm{NN}}}}
\newcommand{\sqrts}{\ensuremath{\sqrt{\mathrm{s}}}}
\newcommand{\rr}{\ensuremath{R}}

\newcommand{\gev}{\ensuremath{\mathrm{GeV/}c}}

\newcommand{\antikT}{anti-\ensuremath{k_\mathrm{T}}}

\newcommand{\pT}{\ensuremath{p_\mathrm{T}}}

\newcommand{\pTjetch}{\ensuremath{p_\mathrm{T,jet}^\mathrm{ch}}}

\newcommand{\pTtrig}{\ensuremath{p_{\mathrm{T,trig}}}}

\newcommand{\ET}{\ensuremath{E_{\mathrm T}}}

\newcommand{\NtrigX}{\ensuremath{N_{\rm{trig}}}}

\newcommand{\Njets}{\ensuremath{N_{\rm{jets}}}}

\newcommand{\Drecoil}{\ensuremath{\Delta_\mathrm{recoil}}}
\newcommand{\cRef}{\ensuremath{c_\mathrm{Ref}}}

\newcommand{\pTreco}{\ensuremath{p_\mathrm{T,jet}^\mathrm{ch,reco}}}

\newcommand{\dphi}{\ensuremath{\Delta\varphi}}

\newcommand{\TTSig}{\ensuremath{\mathrm{TT}_{\mathrm{Sig}}}}
\newcommand{\TTRef}{\ensuremath{\mathrm{TT}_{\mathrm{Ref}}}}

\newcommand{\TT}[1]{\ensuremath{\mathrm{TT}}\{{#1}\}}